\documentclass[pra,twocolumn,floatfix,preprintnumbers,superscriptaddress]{revtex4-1}
\usepackage[english]{babel}
\usepackage[latin1]{inputenc}
\usepackage{enumerate}
\usepackage{amsfonts}
\usepackage{amssymb}
\usepackage{amsmath,amsbsy}
\usepackage{dcolumn}
\usepackage{algpseudocode}
\usepackage{bm}
\usepackage{graphicx, graphics}
\usepackage{color}
\usepackage[breaklinks=true, pdftex]{hyperref}
\usepackage{multirow}

\usepackage{multibib}

\hypersetup{
    colorlinks=true,
    linkcolor=[rgb]{0.0,0.2,0.8},
    citecolor=[rgb]{1.0,0.0,0.0},
    filecolor=magenta,
    urlcolor=[rgb]{0.0,0.3,0.7}
}
\newcommand{\pac}[1]{ \left\{ #1 \right\} }
\newcommand{\pap}[1]{\left( #1 \right)}
\newcommand{\pas}[1]{\left[#1 \right]}

\newcommand{\Ho}{{H}}

\newcommand{\rhoo}{{\rho}}
\newcommand{\bra}[1]{\left\langle #1 \right\vert}

\newcommand{\ket}[1]{\left\vert #1 \right\rangle}
\newcommand{\tr}[1]{\mathrm{tr}\left\{ #1 \right\}}
\newcommand{\ii}{\dot{\iota}}
\newcommand{\dd}{\mathrm{d}}

\newcommand{\ups}{\upsilon}

\begin{document}
	\title{Pulsed Generation of Quantum Coherences and Non-classicality in Light-Matter Systems}
	\author{F.~J. G\'omez-Ruiz}
		\email{ fj.gomez34@uniandes.edu.co}
		\affiliation{Departamento de F{\'i}sica, Universidad de los Andes, A.A. 4976, Bogot\'a D. C., Colombia.}
	\author{ O.~L. Acevedo}
		\affiliation{JILA, University of Colorado, Boulder, Colorado, U.S.A.}
	\author{F. J. Rodr{\'i}guez}
		\affiliation{Departamento de F{\'i}sica, Universidad de los Andes, A.A. 4976, Bogot\'a D. C., Colombia.}
        \author{L. Quiroga}
        		\affiliation{Departamento de F{\'i}sica, Universidad de los Andes, A.A. 4976, Bogot\'a D. C., Colombia.}
	 \author{N.~F. Johnson}
        		\affiliation{Department of Physics,  George Washington University, Washington D.C. 20052, U.S.A.}	
	\date{\today}
\begin{abstract}
We show that a pulsed stimulus can be used to generate many-body quantum coherences in light-matter systems of general size. Specifically, we calculate the exact time-evolution of an $N$ qubit system coupled to a global boson field, in response to an up-down pulse. The pulse is chosen so that the system dynamically crosses the system's quantum phase transition on both the up and down portion of the cycle. We identify a novel form of dynamically-driven quantum coherence emerging 
 for general $N$ and without having to access the empirically challenging strong-coupling regime. Its properties depend on the {\em speed} of the changes in the stimulus. Non-classicalities arise within each subsystem that have eluded previous analyses. Our findings show robustness to losses and noise, and have potential functional implications at the systems level for a variety of nanosystems, including collections of $N$ atoms, molecules, spins, or superconducting qubits in cavities -- and possibly even vibration-enhanced light-harvesting processes in macromolecules.		
\end{abstract}
\maketitle

\section{Introduction}
\begin{figure*}[t]
\begin{center}
\includegraphics[scale=0.27]{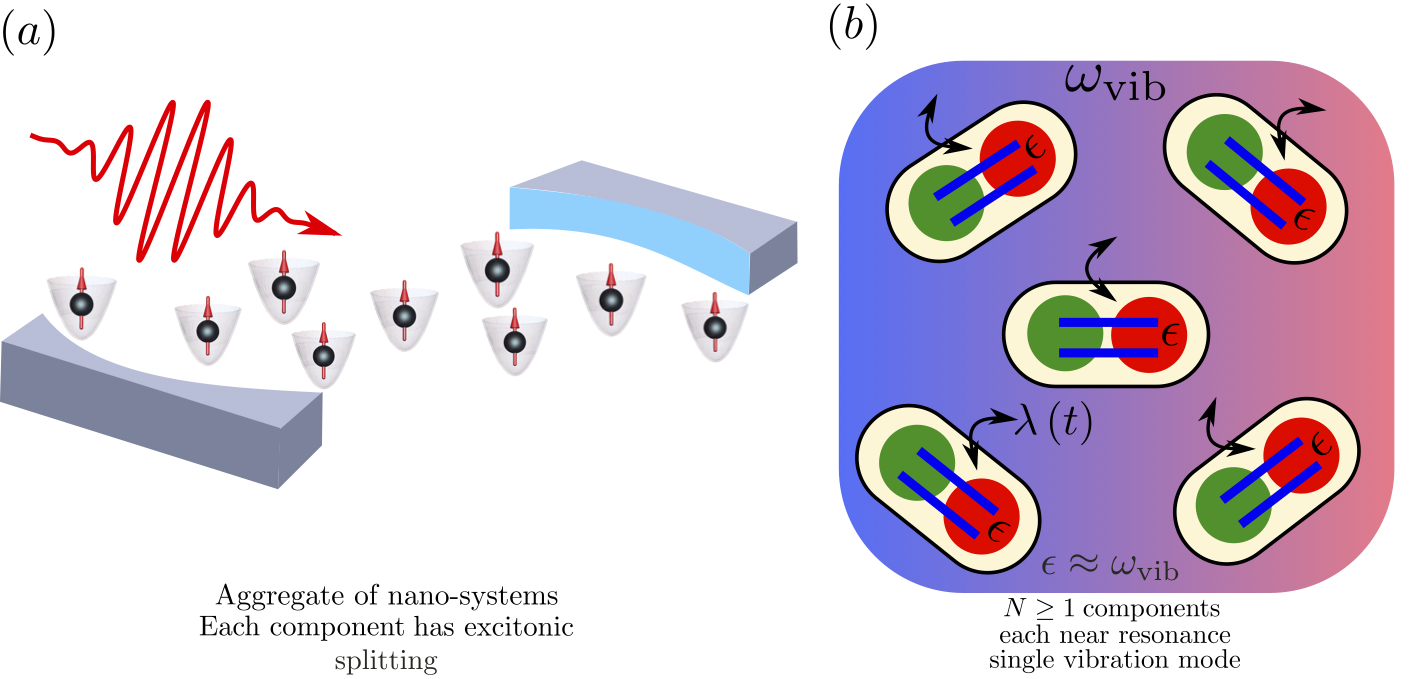}
\caption{(A) {\bf Prototype system.} As an illustration of how our general model and results might be applied in the future, this schematic diagram indicates the type of system that could mimic the dynamics that we analyze for $N$ qubits immersed in a single-mode bosonic environment. We stress that similar implementations have already been built experimentally within the atomic physics community. (B) Schematic representation of the single mode, resonant version of our model (Eq.~\eqref{hdic}). Though it is not our intention to accurately describe any one implementation, we note that in the possible setting of light harvesting/processing  in biochemical systems, each qubit or dimer in Fig.~\ref{fig_1}(A) comprises two split excitonic energy levels with separation ${\epsilon}$ which can be regarded as the basic two-level component in our $N$-component system. The coupling ($\lambda(t)$) is time-dependent in order to capture the complex swathe of additional non-equilibrium, anharmonic interactions that can be generated in the system by an external pulsed stimulus.}\label{fig_1}
\end{center}
\end{figure*}

The interactions between electronic excitations in matter and quantized collective excitations, lie at the heart of conventional condensed matter physics - in which the focus is on periodic systems - as well as nanostructures which are increasingly being fabricated from materials of common interest to chemists, physicists and biologists. Characterizing how collective quantum behavior can be generated in such systems, and what its properties are, represents a challenging research area -- and also an important technological one, e.g. for quantum information processing -- since each system is ultimately a many-body quantum system embedded in an environment. Of particular interest is the issue of correlations or `coherence' in such systems, which in its purest quantum mechanical form manifests itself as many-body quantum entanglement. Recently, new experimental setups have shown a high degree of control of coherence in scenarios involving elementary boson excitations or confined photons interacting with atoms, molecules or artificial nanostructures in cavities~\cite{Yuan2018,Yuan2017,castroPRL}. Interest in the resulting collective coherences now extends beyond the realm of inorganic materials, to organic and biomolecular systems for which there is an ongoing debate concerning the origin and robustness of such quantum coherences in warm environments ~\cite{1,Yuan2017,Flick}. For example, the recent {\em Nature} review of Scholes et al.~\cite{1} tentatively points toward a surprising ubiquity of coherence phenomena across chemical and biophysical systems that are driven by some external stimulus -- typically a high-power light source which provides time-dependent perturbations that generate vibrational responses on the ultrafast scale~\cite{Tsakmakidiseaaq,1,4,5,Jha1,8,23,33,41}.  It is suspected that many of these coherence phenomena involve some generic form of  quantum mechanical interference between the many-body wave function amplitudes of the system's electronic and vibrational (i.e. boson field) components~\cite{1,5}. Indeed there is a body of evidence~\cite{1,4,5,8,23,33,41,Stockman} suggesting that coherence  phenomena in chemical and biophysical systems of general size can show a surprising level of robustness and extended survival time in the presence of noise. Reference~\cite{1} suggests that these observations are so ubiquitous that focus should be turned toward exploring the connection between coherence and possible biological function. \\
\\
Unfortunately, it is impossible to evaluate the exact quantum evolution of a driven mixed exciton-carrier-vibrational system of arbitrary size. Any theoretical analysis will therefore, by necessity, make approximations in terms of the choice of specific simplifying geometries, the specific number of system components included in the calculation (e.g. $N=1$ dimer as in Ref.~\cite{5}), choices about the coupling between the various excitations of the system, and the manner in which memory effects are averaged over or truncated. While convenient computationally, such approximations have left open the question of the fundamental nature of such coherence phenomena, and how they might possibly be generated as the number $N$ of system components increases towards the tens, hundreds or thousands as in real experimental samples. This highlights the need for calculations that purposely avoid these conventional approximations, albeit while making others, in an effort to better understand the general many-body problem for arbitrary $N$ and arbitrary matter-boson coupling strength.\\
\\
Here we study how many-body quantum coherences (specifically quantum entanglement) can be generated, and perhaps ultimately understood, for a rather generic nanostructure system coupled to a bosonic filed and subject to an external stimulus. Our approach to capturing the effects of a time-dependent field-matter interaction is via the modulation of the matter polarization generated by a time-dependent, externally applied pulse stimulus. While our calculations are not specifically designed to mimic any particular physical nor biochemical nanostructure system, we illustrate our results by referring to a hybrid qubit matter system coupled to a single-mode boson field. While we freely admit that our calculations lack the fine details of other works targeted at specific experimental systems, the generic nature of our calculations allows an examination of what might currently be missed from other calculations that adopt the traditional approximations. Our calculations predict that strong quantum coherences and non-classicalities can be generated surprisingly easily in such a driven nanosystem comprising a general number $N$ of components (Fig.~\ref{fig_1}) without needing to access the strong-coupling regime, but instead as a result of the internal dynamics -- in particular, the {\em speed} of the changes. As a corollary, our findings suggest that strong quantum coherences will already have been generated in experiments to date that happen to have fallen in this broad speed regime, and hence offer a possible unified explanation of these. While of course not approximation-free, our theoretical approach avoids the most common approximations listed above, and yields results that in principle apply to any number of components $N$, and include memory effects directly. The Hamiltonian that we consider is purposely simpler and more generic than many studied to date for typical radiation-matter systems. Specifically, we consider a time-dependent generalization of the Dicke model (DM)~\cite{Dicke_PR}, noting that the static DM has a second-order quantum phase transition in the thermodynamic limit of infinite $N$. We focus on understanding the conditions under which large quantum coherence and non-classicality  are generated. As a result, our findings may help enhance understanding of the potential functional advantages that such collective quantum coherences offer at the level of an entire system for general $N$.
In the following sections, we first provide a detailed justification for our approach and its general applicability. We then present our main quantitative results before discussing the overall robustness in the presence of noise and losses.

\section{Methods}
\subsection{Driven system of arbitrary size}
In the past few decades, the development of new experimental techniques for quantum control has led to important advances in the characterization of light-matter systems. Given that no perfect model exists, and that added realism rapidly makes a calculation for general $N$ intractable, we choose a minimal model that is generic enough to broadly mimic different experimental setups, yet is not specific enough that it is weighed down by myriad chemical, biological or physical details. This of course comprises its direct applicability to any specific experimental system, however by so doing it allows us to focus on the resulting quantum coherences in a way that previous calculations could not. Specifically, we consider an arbitrary number $N$ of nanosystem components (e.g. $N$ identical qubits/dimers from Fig.~\ref{fig_1}) whose excitonic levels become coupled to a particular bosonic (e.g. vibrational) mode of the system, as in  Fig.~\ref{fig_1}(b). The coupling is enhanced by dynamical fields that can be created inside the system as a result of a strong external stimulus (e.g. pulse of light). The driven system comprises a complex mix of time-dependent interactions which might be modeled either by anharmonic terms or -- in the simplest way -- by adopting an effective time-dependent matter-boson field coupling, as we do here. Reference \cite{PRBluis} further demonstrates the reasonability of this approximation for the explicit case of control of non-Markovian effects in the dynamics of polaritons generated in semiconductor microcavities at high laser-pumping pulse intensities. As a result, an effective classical intensity sets the coupling strength which becomes time-dependent. The resulting Hamiltonian can then be represented in highly simplified form as a time-dependent generalized Dicke-like model for any $N\geq 1$ \cite{Acevedo2014PRL}
\begin{equation}\label{hdic}
\begin{split}
{H_N}(t)=&\sum_\beta\omega_\beta {a_\beta}^{\dagger}{a_\beta} + \sum_{i=1}^{N}\sum_{\alpha_i\in i}\frac{\epsilon_{\alpha_i}}{2}{\sigma}_{z,\alpha_i}^{i} \\
&+\sum_\beta  \sum_{i=1}^{N}\sum_{\alpha_i\in i}
\frac{\lambda_{\alpha_i,\beta}^{i}(t)}{\sqrt{N}}\left({a_\beta}^{\dagger}+{a_\beta}\right){\sigma}_{x,\alpha_i}^{i}
\end{split}
\end{equation}
where ${\sigma}_{p,\alpha_i}^{i}$ denotes the Pauli operators for excitation $\alpha_i$ on each component (e.g. qubit/dimer, Fig.~\ref{fig_1}(b)) $i$ with $p=x,z$. The first term is the set of boson/vibrational modes $\{\beta\}$ which in general may or may not be localized around certain locations. The second term represents the qubit excitations $\{\alpha_{i}\}$ localized on each of the components  $i=1,\dots,N$ (e.g. qubit/dimers, Fig.~\ref{fig_1}(b)). For instance, the two electronic states on each component may be hybrid excitonic states, e.g. $|X\rangle$ and $|Y\rangle$, in a dimer. The third term gives the coupling between the electronic and bosonic (e.g. vibrational) terms, by means of which energy and quantum coherence can be transferred back and forth between these matter components $\{\alpha_{i}\}$ and the bosonic modes $\{\beta\}$. We stress that our choice of $N$ components in Eq.~\eqref{hdic} does not mean that this is necessarily the total number of qubit-like units in the system under study: It may happen that in practice only some portion of the system is probed by the experiment, hence $N$ can in principle be tailored to account for this.

Equation~\eqref{hdic} is quite general in terms of its scalability to any number of components, and can serve a similar function to models such as the Ising model in getting at the general universality of behaviors to be expected across materials \cite{Acevedo2014PRL}. This is important given the wide range of chemically and physically diverse systems in which generic coherence effects are observed~\cite{1,4,5,8,23,33,41} which in turn motivates our generic as opposed to material-specific approach. We have already shown that for variants of Eq.~\eqref{hdic} there is a universal dynamical scaling behavior for a particular class concerning their near-adiabatic behavior, in particular the Transverse-Field Ising model, the Dicke Model and the Lipkin-Meshkov-Glick model~\cite{Acevedo2014PRL}. Since the entire system is for the moment considered closed, there is no overall dissipation but it does allow for the fact that the molecular subsystem components may be losing energy to the bosonic (e.g. vibrational) subsystem and vice versa. It also makes no assumptions about the memory in either the molecular dynamics or any vibrational system, for example, or their coupling. Specifically, it is non-Markovian by design; it includes all memory effects; it is valid irrespective of how fast or slow $\lambda_{\alpha_i,\beta}^{i}(t)$ varies or its temporal profile; it applies irrespective of the individual spectra at each site of the spectrum of bosonic modes; and most importantly, it applies to any value of $N$.

Our focus here is on near resonant conditions since these are the most favorable for generating large coherences. Hence we assume for the moment that each component $i$ has one multi-electron energy level separation that is approximately the same as one of the possible bosonic (e.g. vibrational) energies, and is also approximately the same for all $N$ components. All other electronic excitation and bosonic modes will be off resonance: including them would modify the quantitative values in our results but the main qualitative findings would remain. Figures~\ref{fig_1}(a) and~\ref{fig_1}(b) provide a motivation for the components in our model, inspired by real systems of atoms/molecules/spins in cavities or dimers in biochemical vibrational environments, and a schematic of the resonant version of our model (Eq.~\eqref{hdic}) comprising $N$ qubits or dimer pairs, where each has two electronic states which are energy-split by $\epsilon$. Such systems have already been successfully controlled in experiments such as cavities containing {\rm GaAs} semiconductor quantum rings~\cite{castroPRL}; single azulene molecules~\cite{Flick}; a chain of ${\rm Na}_2$ dimers~\cite{Flick2}; and also organic systems~\cite{Orgiu} coupled to molecular resonators with a microcavity mode~\cite{shala} or Raman scattering~\cite{pino}. All these systems are broadly consistent with a model of qubits comprising two hybridized excitonic states with energy splitting $\epsilon=\sqrt{\Delta^2+4V^2}$, where $\Delta$ is the energy exciton splitting and $V$ is the direct dipole-dipole coupling strength, with the whole system immersed in a vibrational single mode cavity or effective environment. We also note that even this single-mode resonance assumption can be generalized by matching up different  excitation energies $\epsilon'$, $\epsilon''$, etc. to the nearest vibrational energies $\omega'$, $\omega''$ etc. and then solving Eq.~\eqref{hdic} in the same way for each subset $(\epsilon',\omega')$ etc. For example, if the $N$ components are partitioned into $n$ subpopulations, where each subpopulation has the same resonant energy and vibrational mode but where these values differ between subpopulations, the total Hamiltonian will approximately  decouple into $H^{(1)}\oplus H^{(2)}\oplus H^{(3)}\dots \oplus H^{(N)}$. Any residual coupling between these subpopulations might then be treated as noise, as discussed later.
%\newpage
%========================= Results and discussions
\section{Results and discussions}
We adopt the simplest single boson mode case which then corresponds to the celebrated Dicke model comprising a set of $N$ identical qubits/dimers symmetrically coupled to a single-mode boson field. It can be described by the microscopic Hamiltonian
\begin{equation}
H\pap{t}=\frac{\epsilon}{2}\sum_{i=1}^{N}\sigma_{z}^{i}+\omega \hat{a}^{\dagger}\hat{a} + \frac{\lambda\pap{t}}{2\sqrt{N}}\sum_{i=1}^{N}\left(\hat{a}+\hat{a}^{\dagger}\right)\hat{\sigma}_{x}^{i}
\end{equation}
The energies $\epsilon$ and $\omega$ represent the qubit/dimer splitting and radiation/vibronic quantum respectively, while $\lambda\pap{t}$ represents the time-dependent interaction. With a single resonance across all $N$ components in Eq.~\eqref{hdic}, and all $N$ components having the same resonant excitation energy, the entire $2^{\otimes N}\otimes \mathbb{N}$ dimensional Hamiltonian reduces down to $SU(2)$ collective operators ${J}_{\alpha}=\frac{1}{2}\sum_{i=1}^{N}{\sigma}^{i}_{\alpha}$ where we now drop all the unnecessary component indices. Equation \eqref{hdic} then reduces exactly to $\hat{H}=\epsilon {J}_{z} + \omega {a}^{\dagger}{a} +\frac{2\lambda(t)}{\sqrt{N}}{J}_{x}\left({a}^{\dagger}+{a}\right)$. For a completely {\em static} coupling $\lambda$ and in the limit $N\to\infty$, there is an electronic-bosonic system phase-boundary at $\lambda_{c} =\frac{\sqrt{\epsilon\omega}}{2}$. For all the results we discuss below, the state at $t=0$ is a direct product of the excitonic and boson field states. The coupling $\lambda\pap{t}$ is turned on smoothly from $\lambda(t=0)=0$ following an up-and-down cycle taken as a triangular shaped pulse. For time-dependent coupling $\lambda(t)$ and finite system size $N$, this ideal phase transition is not completely achieved -- however its precursors are what generate the new forms of collective quantum coherence and non-classicality presented in Figs.~\ref{fig2} and~\ref{fig3}, respectively.

\subsection{Speed drives multi-component quantum coherence}
Since we are interested in the system's quantum coherences and non-classicality following pulsed perturbations, we take $\lambda(t)$ to be a piecewise linear ramping up and down for simplicity, i.e. triangular profile with total round-trip time $\tau$ which acts as an inverse annealing velocity (ramping velocity) $\upsilon$  and for the single resonant condition $\epsilon=\omega=1$. The precise details of  $\lambda(t)$ in any particular experiment will depend on the type of nonlinearities induced by the particular probing method, but similar qualitative features in our results will appear for any up-and-down form.

\begin{figure*}[t]
\begin{center}
\includegraphics[scale=0.35]{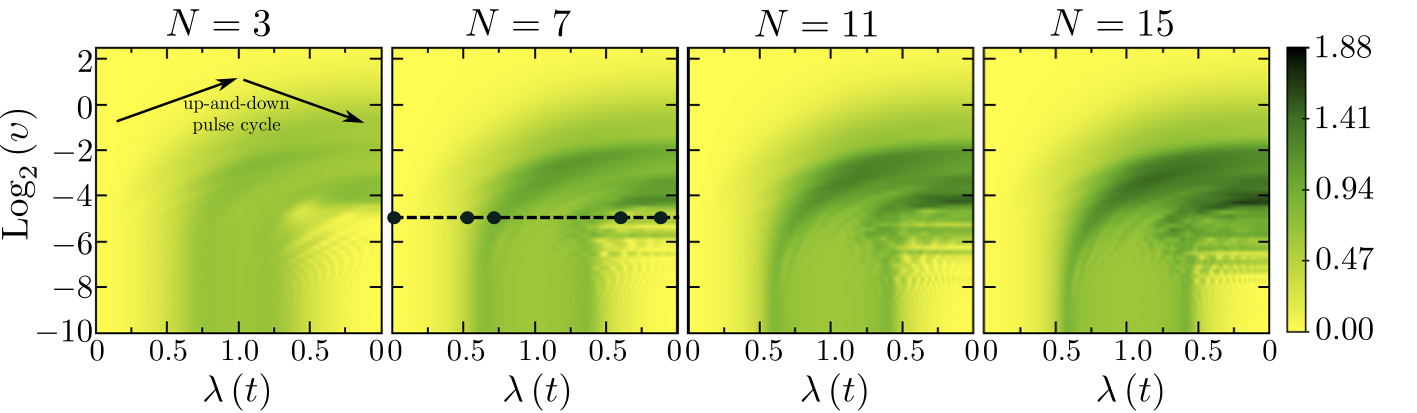}
\caption{Collective quantum coherence generated by a simple up-and-down pulse (i.e. triangular $\lambda(t)$ indicated in first panel), as measured by the von Neumann entropy which quantifies the quantum entanglement between the electronic and bosonic (e.g. vibrational) subsystems. By the end of just one up-and-down cycle for a broad range of intermediate return trip times, a substantial amount of quantum coherence is generated in the $N$-component system for general $N$. If the external perturbation is then turned off, for example because the pulse has ended, the generated coherence will survive as long as the built-in decoherence/dephasing mechanisms in the sample allow it to last. The darker the color, the larger the quantum coherence (see color bar).  The larger the $\upsilon$, the less negative the logarithm (i.e. higher on the vertical scale), and the shorter the return trip time. Since these results look qualitatively similar for any $N\geq 3$, they offer insight into the ubiquity of coherences observed empirically in chemical and biophysical systems \cite{1}. Increasing $N$ simply increases the numerical value of the peak value, while choosing a smaller $\lambda(t)$ maximum just reduces the magnitude of the effect. The five points indicated along the horizontal dashed line for $N=7$, correspond to the five specific values of time at which the sub-system Wigner functions are evaluated in the next section. }\label{fig2}
\end{center}
\end{figure*}

\begin{figure*}[t]
\begin{center}
\includegraphics[scale=0.38]{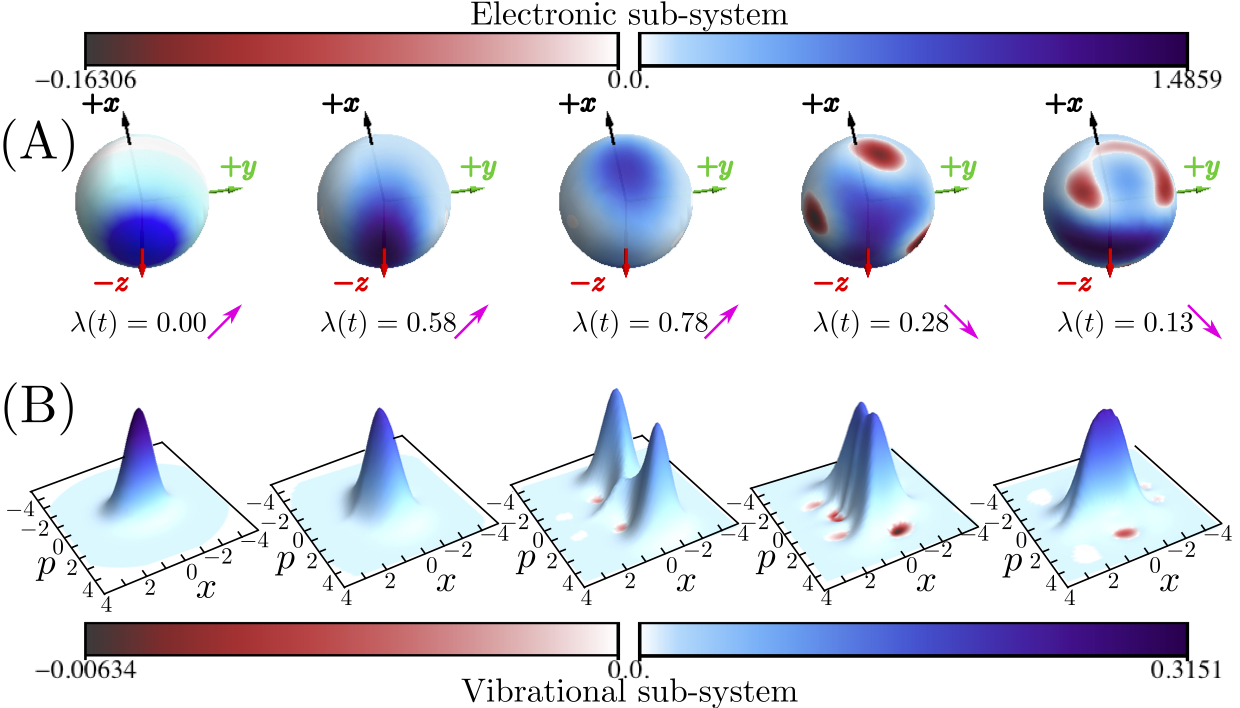}
\caption{(A) Electronic sub-system Agarwal-Wigner Functions $W_q$ and (B) boson field/vibrational sub-system Wigner functions $W_b$, shown at two values of $\lambda(t)$ in each portion of an up-and-down pulse cycle. The pulse cycle is depicted as the horizontal dashed line in Fig.~\ref{fig2} for $N=7$. The $\upsilon$ value is purposely chosen not to be the optimal one producing the strongest coherence $\pap{\log_{2}\pap{\upsilon}=-5}$, because we want to illustrate the type of non-classicality that can be achieved for broader values of $\upsilon$. Most importantly, by the end of just one up-and-down cycle, both $W_q$ and $W_b$ develop complex non-classical patterns for a broad range of intermediate return trip times and general $N\geq 3$ (see Fig.~\ref{fig2}). $W_q$ and $W_b$ are phase space representations. Though positive portions may be quantum mechanical or classical, the negative portions (red and black) that appear demonstrate unambiguous non-classicality. In (A), opposite Bloch hemispheres are not shown because of symmetry: $W_q(\theta,\phi+\pi)=W_q(\theta,\phi)$. In (B), $W_b$ is represented in the $x-p$ plane of position (vertical) and momentum (horizontal) quadrature.}\label{fig3}
 \end{center}
 \end{figure*}

We consider ramping up to $\lambda(t)\approx 1$ and back, though we stress that similar (but weaker) features will be seen for smaller maximum values. For each time $t$ starting at $t=0$, we obtain numerically the instantaneous state $\ket{\psi(t)}$. Since we are interested in the {\em additional} coherence generated by the dynamics, we start at $t=0$ with $\ket{\psi(0)}=\bigotimes_{i=1}^{N}\ket{\downarrow}\otimes\ket{n=0}$ where both electronic and bosonic subsystems have zero induced excitations. Again, this can be generalized without changing the main details. The accuracy of our numerical solutions was checked by extending the expansion basis beyond the point of convergence. For general ramping velocity $\upsilon$, the amplitude of being either in the ground or the collected excited states, accumulates a dynamical phase with these  channels interfering with each other and hence forming the oscillatory patterns.  At low ramping velocities, the near-adiabatic regime has a general tendency to show an increase in memory effects as the cycles get faster. However for a broad range of intermediate ramping velocities (Fig.~\ref{fig2}) a new regime emerges which is characterized by large quantum coherence between the bosonic (e.g. vibrational) and electronic subsystems. This process would represent a squeezing mechanism in both the electronic and vibrational subsystems, followed by the generation of electronic-vibrational coherence in the form of genuine quantum mechanical  entanglement~\cite{AcevedoPRA2015, Acevedo2015NJP, Gomez_Ent2016}. As the annealing velocity is further increased, the system has less and less time to undergo any changes.

Figure~\ref{fig2} quantifies this electronic-bosonic (e.g. electronic-vibrational) quantum coherence generated by the applied pulse in terms of the entanglement as measured by the von Neumann entropy. Given a subsystem $A$, the von Neumann entropy:
\begin{equation} \label{eqentro}
S_N=-\tr{\rhoo_A\log \pap{\rhoo_A }}\:,\:\:\:\rhoo_A=\mathrm{tr}_B\pac{\ket{\psi}\bra{\psi}}
\end{equation}
where $B$ is the complementary subsystem and the total system is in a total state $\ket{\psi}$ that is pure. When the total system is in such a pure state, the entropy of subsystem $A$ is equal to the entropy of its complementary subsystem $B$, and this quantity $S_N$ is a measure of the entanglement between both subsystems. The natural choice in our system for such a bipartition is where one subsystem is the bosonic (e.g. vibrational) mode and the other subsystem is the matter (e.g. molecular excitonic) subsystem. Since this a closed system (i.e. a pure global quantum state with an unitary evolution), the increase of $S_N$ in each subsystem is synonymous with an interchange of information between the bosonic vibrations and matter components during the cycle, hence providing a more direct thermodynamical interpretation for the memory effects of the cycle.

The collective coherence in Fig.~\ref{fig2} is purely quantum in nature  (i.e. entanglement); it involves an arbitrary number $N$ of components ($N\geq 3$); and it is achieved using any up-and-down $\lambda(t)$ and {\em without} the need to access the strong matter-bosonic field (e.g. electron-vibrational) coupling limit. This is important in practical terms since strong coupling can be hard to generate and control in a reliable way experimentally. Instead, as illustrated in Fig.~\ref{fig2} for each value of $N$, we find that the same macroscopic coherence is generated by choosing intermediate ramping velocities and undergoing a return trip, as shown. Moreover the same qualitative result as Fig.~\ref{fig2} holds for any $N\geq 3$ and becomes stronger with $N$. Hence we have shown that  by the end of just one up-and-down cycle for a broad range of intermediate return trip times, a substantial amount of quantum coherence will have been generated in the $N$-component system for general $N$. This enhanced entanglement region can be seen as bounded by a maximum ramping velocity $\ups_{\mathrm{max}}$ above which the sudden quench approximation is valid, and a minimum ramping velocity $\ups_{\mathrm{min}}$ below which the adiabatic condition is fulfilled. $\ups_{\mathrm{min}}$ does not depend on the maximum value of $\lambda(t)$ reached, which is to be expected since the ground state in the ordered phase has an asymptotic of $S_N \rightarrow \log 2$ and the adiabatic condition should only depend on the system size $N$. The scaling $\ups_{\mathrm{min}} \propto N^{-1}$ that emerges, comes from a relation for the minimal energy gap at the critical threshold~\cite{AcevedoPRA2015}. The upper bound $\ups_{\mathrm{max}}$ does not depend on system size. In the near adiabatic regime, the von Neumann entropy is not always increasing with time, which means that for slow annealing velocities, information is not always dispersing from the vibrational subsystem to the molecular subsystem and vice versa. Instead, there is some level of feedback for each subsystem, so that they are still able to retain some of their initial state independence. However, this feedback becomes increasingly imperfect so that at annealing velocities near the boundary with the intermediate regime, the information mixing attains maximal levels. After that, the mixing of information between vibrational and electronic subsystems is always a monotonic dispersion process, which becomes reduced as the time of interaction is reduced more and more. This establishes a striking difference between the lack of memory effects in the adiabatic and sudden quench regimes: the former's cycle comprises a large but reversible change, while the latter's cycle is akin to a very small but irreversible one. In practice, both mean relatively small changes to the initial condition -- however this is a consequence of two very different properties. This interplay between actual change and its reversibility may explain why the transition between those two regimes is more intricate that might have otherwise been imagined.

\subsection{Multi-component non-classicality}
Our system shows the novel feature of demonstrating non-classicality in both the vibrational {\em and} the electronic subsystems for arbitrary $N$.
Specifically, Fig.~\ref{fig3} shows  this non-classicality generated separately within each subsystem during the up-and-down $\lambda(t)$ cycle, and is represented by the Agarwal-Wigner-Function and Wigner quasi-distributions for the electronic and vibrational subsystems respectively. As $\lambda(t)$ increases from zero, the Wigner Function exhibits squeezing, with the Wigner function then splitting along the $x$ and $-x$ directions and no longer concentrated around the initial state. Increasing $\lambda(t)$ further leads to appearance of negative scars (see red portions) which are uniquely non-classical phenomena -- though we stress that even  positive portions of $W_q$ and $W_b$ can exhibit quantum mechanical character. Both $W_q$ and $W_b$ not only develop multiple negative regions which are a marker of non-classical behavior, but they also contain so-called sub-Planckian structures which have been related to quantum chaos. Most importantly, by the end of just one up-and-down cycle, both $W_q$ and $W_b$ have developed complex non-classical patterns, with a blend of regular and chaotic character.

\subsection{Impact of losses and noise}
\begin{figure*}[t]
\begin{center}
\includegraphics[scale=0.39]{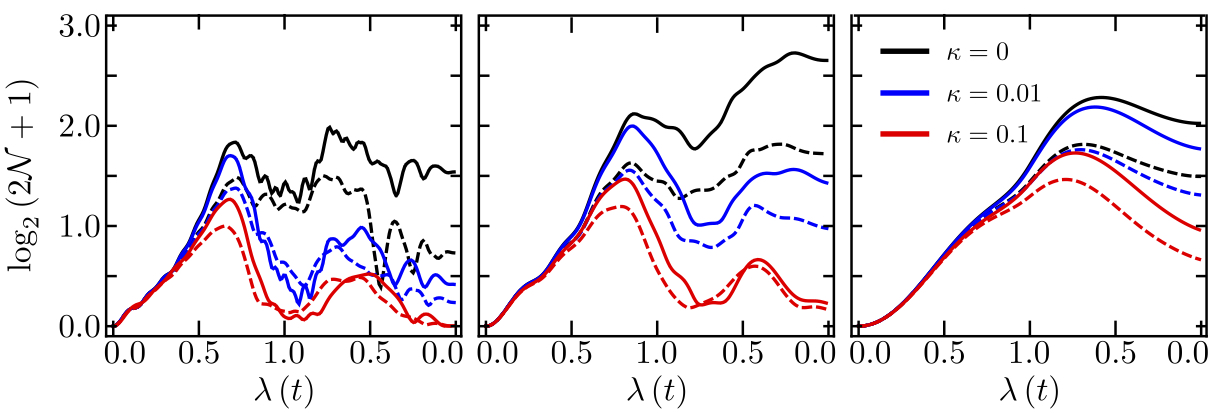}
\caption{We show evidence of the robustness of the many-body electronic-vibrational entanglement, as witnessed by quantum logarithmic negativity, to decoherence/losses. Results are shown for three representative, intermediate duration up-and-down pulses (i.e. the annealing velocities $\upsilon$ for left, middle, right panels are $\log_{2}\pap{\upsilon}=-4.64,\, -3.32,\, -1.32$ respectively).  Results are shown for $N=5$ (dashed lines) and $N=11$ (solid lines) and for several values of decoherence $\kappa$.}\label{fig4}
 \end{center}
 \end{figure*}

Following the density matrix approach of Ref.~\cite{AcevedoPRA2015}, we have investigated numerically how the presence of decoherence/losses to the environment in the chemical or biophysical system will affect the dynamics discussed above, as illustrated in Fig.~\ref{fig4}. The widely-accepted best entanglement measurement in an open quantum system is the quantum negativity $\mathcal{N}\left(\rho\right)=\frac{1}{2}\left(\left| \rho^{\Gamma}_{q}\right|_{1} - 1 \right)$ where $ \rho^{\Gamma}_{q}$ is the partial transpose of $\rho$ with respect to the electronic subsystem, and $\left|\hat{\mathcal{O}}\right|_{1}\equiv \mathrm{tr}\pac{\sqrt{\hat{\mathcal{O}}^{\dagger}\hat{\mathcal{O}} }} $  is the trace norm. The electronic-vibrational density matrix $\rho\pap{t}$ evolves as~\cite{breuer}:
\begin{equation}
\frac{\dd}{\dd t}\hat{\rho} = -\ii \pas{\Ho,\hat{\rho}} +2\kappa \pap{\bar{n}+1} \mathcal{L}\pap{\hat{\rho};\hat{a}}+2\kappa \bar{n} \mathcal{L}\pap{\hat{\rho};\hat{a}^{\dagger}}
\label{eqME}
\end{equation}
where the Lindblad superoperator $\mathcal{L}\pap{\rho;\hat{\mathcal{O}}}$ for the arbitrary operator $\hat{\mathcal{O}}$ is defined as $\hat{\mathcal{O}}\rho\,\hat{\mathcal{O}}^{\dagger}-\frac{1}{2}\pac{\hat{\mathcal{O}}^{\dagger}\hat{\mathcal{O}},\rho}$ and $\pac{\bullet,\bullet}$ is the traditional anti-commutator. Moreover, $\kappa$ is the damping rate and $\bar{n}$ is the thermal mean photon number.
All our main results survive well if the decoherence term through interaction with the environment, is anywhere up two orders of magnitudes lower than the main energy scale. Furthermore, even if dissipation is at values of just an order of magnitude below, spin squeezing effects remain highly robust, with increasing noise resistance with system size. Vibrational field squeezing surprisingly survives to dissipation regimes comparable to the Hamiltonian dynamics itself. On the other hand, detailed features of the chaotic stage (such as order parameter oscillations, negative regions, and sub-Planck structures) are far more sensitive to decoherence. These very sensitive features could be used as tools for measuring very weak forces. In our analysis, we have found that introducing small but finite values of the average number of phonons $\bar{n}$ (such as those typical at the ultra-low temperatures in most experimental realizations) does not change qualitatively the conclusions; it just slightly intensifies the process of decoherence.

We have so far assumed a single $\epsilon,\omega$ pair are close to each other in energy. In the limit that other pairs are also near resonance but these resonances have very different energies from $\epsilon,\omega$, a similar dynamical coherence can develop within each of these subspaces of the full Hamiltonian (Eq.~\eqref{hdic}). Each pair will have its own up-and-down return trip time (and hence ramping velocity $\upsilon$) for which the coherence is maximal. Since the full Hamiltonian can then be written approximately as a sum of these separate subspaces, the full many-body wave function will include a product of the coherent wavefunctions $\Psi_{\epsilon',\omega'}(t)$ for these separate $\{\epsilon',\omega'\}$ subspaces. In the more complex case where several pairs are close together in energy, they will each tend to act as noise for each other. Suppose that the coherence for pair $\epsilon,\omega$ is described by $\Psi_{\epsilon,\omega}(t)$ and it is perturbed by noise from two pairs $\{\epsilon'',\omega''\}$ and $\{\epsilon''',\omega'''\}$ which happen to be nearby in energy. The fact that they are dynamically generated in the same overall system due to the same incident pulses, means that they will likely represent correlated noise. Such correlated noise from various sources can actually help maintain the coherence of $\Psi_{\epsilon,\omega}(t)$ over time. To show this, consider the following simple example (though we stress that there are an infinite number of other possibilities using other numbers and setups, see Ref.~\cite{Lee}) in which we treat $\Psi_{\epsilon,\omega}(t)$ for the pair $\epsilon,\omega$ as a two-level system. The two subspaces $\{\epsilon'',\omega''\}$ and $\{\epsilon''',\omega'''\}$ each generate decoherence of $\Psi_{\epsilon,\omega}(t)$ in the form of discrete stochastic phase-damping kicks. Such phase kicks are a purely quantum mechanical mechanism for losing coherence, as opposed to dissipation.
The probability distributions of the kicks from these two subspaces are $P_A, P_B$. In addition, the kicks are such that the kick of $\Psi_{\epsilon,\omega}(t)$, described by the rotation angle $\theta_2$ is correlated to the previous rotation angle ($\theta_1$):
\begin{widetext}
\begin{equation}
\begin{split}
P_A(\theta_2|\theta_1)&=\begin{cases}
\frac{1}{3}[ \delta(\theta_2) + \delta(\theta_2+\frac{\pi}{2}) + \delta(\theta_2-\frac{\pi}{2})], &\qquad\text{if $\theta_1 \in \{-\frac{\pi}{2}, 0, \frac{\pi}{2} \}$,}\\
\delta(\theta_2), & \qquad\text{otherwise}
\end{cases}\\
 P_B(\theta_2|\theta_1) &=\begin{cases}
 \frac{1}{3}[ \delta(\theta_2-\epsilon)  +\delta(\theta_2+\frac{3 \pi}{4}) + \delta(\theta_2-\frac{\pi}{4} ) ],&\text{if $\theta_1 \in \{-\frac{3 \pi}{4}, \epsilon, \frac{\pi}{4} \}$}\\
 \delta(\theta_2-\epsilon),&\text{otherwise}
 \end{cases}
\end{split}
\end{equation}
\end{widetext}
with similar conditions holding for all subsequent pairs $\theta_i$ and $\theta_{i-1}$ (see Ref. \cite{Lee} for general discussion). The specific choice of angles may be generalized. The parameter $\epsilon$ is small, and its presence just acts as a memory of which probability distribution was selected in the previous step. If $P_A$ represents the only noise-source applied, and assuming the initial angle of rotation is $0$ (i.e. $\theta_1=0$) then $P_A(\theta_n, \ldots, \theta_1)= \prod_{i=2}^{n} P_A(\theta_{i}| \theta_{i-1})= (\frac{1}{3})^{n-1}$. Hence if under the influence of subspace $\{\epsilon'',\omega''\}$ (and hence $P_A$), the density matrix for $\Psi_{\epsilon,\omega}(t)$ will have off-diagonal elements (which correspond to the decoherence) that decrease by a factor $\frac{1}{3}$ after each phase-kick.
Similar arguments hold if $P_B$ is the only noise-source applied to the system and if we assume $\theta_1 =\epsilon$. Combining the two noise-sources (i.e. probability distributions) at random means that the angles of rotation can take on seven values, $\{ -\pi/3 , -\pi/2, 0,  \epsilon, \pi/3, \pi/2, \pi \}$. The decay factor now becomes exactly $2/3$ in the  limit of $\epsilon\rightarrow 0$. This means that the {\em combination} of the noise sources causes a slower decoherence of $\Psi_{\epsilon,\omega}(t)$ than each on their own. Hence it is  possible that the quantum coherence of $\Psi_{\epsilon,\omega}(t)$ due to a near resonance of $\epsilon,\omega$ as studied in detail in this paper (Figs.~\ref{fig2}-\ref{fig2}) is actually favored by having competing coherence processes in the same system.
%\newpage
\section{Conclusions}
Our results show that nanosystems of fairly general size and driven by pulses (e.g. due to a high power external light source or some other applied field) can show surprisingly strong quantum coherence and non-classicality without necessarily passing to the strong coupling regime, but instead through its dynamics -- in particular, the {\em speed} of the dynamical changes that are induced. As we show in Fig.~\ref{fig2}, the resulting coherence builds up during the up-and-down ramping associated with an external driving pulse (e.g. light pulse) and is large at the end of it. If this ramping is then turned off, for example because the pulse has ended, the generated coherence will survive as long as the built-in decoherence/dephasing mechanisms allow it to last. Our calculations show that it could remain for a significant time if the noise is not too large. Our approach complements existing work in that we avoid the usual type of approximations prevalent in the quantum coherence literature~\cite{1} and instead presents results that in principle apply to general $N\geq 3$. The Hamiltonian that we consider is purposely simpler and more generic than many studied to date in order that we can focus attention on understanding the conditions under which optimal coherence can be generated and hence become available for functional use. Though we considered the coupling $\lambda$ to be taken to a relatively modest value ($\sim 1$) and returned, even lower maximum values will give qualitatively similar effects.

Among natural or artificial nanosystems for which these findings could be relevant, we make specific mention of aggregates of real or artificial atoms in cavities and superconducting qubits~\cite{Ciuti2010,Marquadt2011}, as well as trapped ultra-cold atomic systems~\cite{bloch2012nat,schneider2012rpp,georgescu2014rmp}. Our findings also add to current efforts surrounding the collective generation and propagation of entanglement~\cite{amico2002nature, wu2004prl, RomeraPLA2013, Reslen2005epl, Juan2010pra, Acevedo2015NJP, AcevedoPRA2015}, the development of spatial and temporal quantum correlations~\cite{sun2014pra,FernandoPRB2016}, critical universality~\cite{Acevedo2014PRL}, and finite-size scalability~\cite{Vidal2006, CastanoPRA2011, CastanoPRA2012}. In addition, the effects described in this work may already be accessible under current experimental realizations in a broad class of systems of interest to physicists. As a result, our findings should be of interest for quantum control protocols which are in turn of interest in quantum metrology, quantum simulations, quantum computation, and quantum information processing~\cite{Gernot2006,Rey2007,Dziarmaga2014AP,Hardal_CRP2015,NiedenzuPRE2015}.

\section*{Disclosure/Conflict-of-Interest Statement}

The authors declare that the research was conducted in the absence of any commercial or financial relationships that could be construed as a potential conflict of interest.

\begin{acknowledgments}
N.F.J. is very grateful to He Wang for stimulating discussions. F.J.G-R., F.J.R., and L.Q. acknowledge financial support from Facultad de Ciencias through UniAndes-2015 project \emph{Quantum control of nonequilibrium hybrid systems-Part II}. F.J.G-R and F.J.R acknowledges financial support from Facultad de Ciencias 2018-II. F.J.G-R and F.J.R acknowledges financial support from \emph{Fundaci\'on para la Promoci\'on de la Investigaci\'on y la Tecnolog\'ia} through Banco de la Rep\'ublica project 3646. N.F.J. acknowledges partial support from the National Science Foundation (NSF) under grant CNS 1522693 and the Air Force under AFOSR grant FA9550-16-1-0247. The views and conclusions contained herein are solely those of the authors and do not represent official policies or endorsements by any of the entities named in this paper.
\end{acknowledgments}
\bibliography{mybib2}	
\end{document}